\documentclass[aps,prl,twocolumn,showkeys,reprint,amsmath,amssymb,superscriptaddress,floatfix,footinbib,longbibliography]{revtex4-2}

\usepackage{graphicx}
\usepackage{epstopdf}

\usepackage[T1]{fontenc}
\usepackage[applemac]{inputenc}
\usepackage{lmodern}
\usepackage{amsmath}
\usepackage{amssymb}
\usepackage[english]{babel}
\usepackage{natbib}
\usepackage{ae}
\usepackage{units}

\usepackage{amsmath,amssymb,natbib,bm}
\usepackage{psfrag}
\usepackage{subfigure}
\usepackage{amsthm}

\usepackage{booktabs}

\usepackage{slashed}

\usepackage[americaninductors]{circuitikz}
\usepackage{tikz}
\usetikzlibrary{arrows}
\usepackage{ulem}
\usepackage{color}
\usepackage{url}

\usepackage[colorlinks]{hyperref}
\hypersetup{
	plainpages=true,
	breaklinks=true,
	hypertexnames=false,
	pageanchor=true,
	colorlinks=true,
	linkcolor={blue},
	citecolor={blue},
	urlcolor={blue},
	anchorcolor={black}
}

\usepackage{mleftright} 

\newcommand{\figref}[1]{\mbox{Fig.~\ref{#1}}}
\newcommand{\tabref}[1]{\mbox{Table~\ref{#1}}}

\renewcommand{\eqref}[1]{\mbox{Eq.~(\ref{#1})}}

\newcommand{\figpanel}[2]{Fig.~\hyperref[#1]{\ref*{#1}(#2)}} 
\newcommand{\figpanels}[3]{Fig.~\hyperref[#1]{\ref*{#1}(#2)-(#3)}} 
\newcommand{\figpanelNoPrefix}[2]{\hyperref[#1]{\ref*{#1}(#2)}} 
\newcommand{\figpanelsNoPrefix}[3]{\hyperref[#1]{\ref*{#1}(#2)-(#3)}} 

\newcommand{\ket}[1]{|#1\rangle}

\newcommand{\be}{\begin{equation}}
	\newcommand{\ee}{\end{equation}}
\newcommand{\bea}{\begin{eqnarray}}
	\newcommand{\eea}{\end{eqnarray}}

\begin{document}

\title{Microwave amplification via interfering multi-photon processes \\ in a half-waveguide quantum electrodynamics system}

\author{Fahad Aziz}
\thanks{F.~A, K.~T.~L, and P.~Y.~W contributed equally to this work.}
\affiliation{Department of Physics, National Tsing Hua University, Hsinchu 30013, Taiwan}

\author{Kuan Ting Lin}
\thanks{F.~A, K.~T.~L, and P.~Y.~W contributed equally to this work.}
\affiliation{Department of Physics and CQSE, National Taiwan University, Taipei 10617, Taiwan}

\author{Ping Yi Wen}
\thanks{F.~A, K.~T.~L, and P.~Y.~W contributed equally to this work.}
\affiliation{Department of Physics, National Chung Cheng University, Chiayi 621301, Taiwan}

\author{Samina}
\affiliation{Department of Physics, National Tsing Hua University, Hsinchu 30013, Taiwan}

\author{Yu Chen Lin}
\affiliation{Department of Physics and CQSE, National Taiwan University, Taipei 10617, Taiwan}

\author{Emely Wiegand}
\affiliation{Department of Microtechnology and Nanoscience, Chalmers University of Technology, 412 96 Gothenburg, Sweden}

\author{Ching-Ping Lee}
\affiliation{Department of Physics, National Tsing Hua University, Hsinchu 30013, Taiwan}

\author{Yu-Ting Cheng}
\affiliation{Department of Physics, National Tsing Hua University, Hsinchu 30013, Taiwan}

\author{Ching-Yeh Chen}
\affiliation{Department of Physics, National Tsing Hua University, Hsinchu 30013, Taiwan}

\author{Chin-Hsun Chien}
\affiliation{Department of Physics, National Tsing Hua University, Hsinchu 30013, Taiwan}

\author{Kai-Min Hsieh}
\affiliation{Department of Physics, National Tsing Hua University, Hsinchu 30013, Taiwan}

\author{Yu-Huan Huang}
\affiliation{Department of Physics, National Tsing Hua University, Hsinchu 30013, Taiwan}

\author{Ian Hou}
\affiliation{Institute of Applied Physics and Materials Engineering, University of Macau, Macau}
\affiliation{UMacau Zhuhai Research Institute, Zhuhai, Guangdong, China}

\author{Jeng-Chung Chen}
\affiliation{Department of Physics, National Tsing Hua University, Hsinchu 30013, Taiwan}
\affiliation{Center for Quantum Technology, National Tsing Hua University, Hsinchu 30013, Taiwan}

\author{Yen-Hsiang Lin}
\affiliation{Department of Physics, National Tsing Hua University, Hsinchu 30013, Taiwan}
\affiliation{Center for Quantum Technology, National Tsing Hua University, Hsinchu 30013, Taiwan}

\author{Anton Frisk Kockum}
\affiliation{Department of Microtechnology and Nanoscience, Chalmers University of Technology, 412 96 Gothenburg, Sweden}

\author{Guin Dar Lin}
\affiliation{Department of Physics and CQSE, National Taiwan University, Taipei 10617, Taiwan}
\affiliation{Physics Division, National Center for Theoretical Sciences, Taipei 10617, Taiwan}
\affiliation{Trapped-Ion Quantum Computing Laboratory, Hon Hai Research Institute, Taipei 11492, Taiwan}

\author{Io-Chun Hoi}
\email{iochoi@cityu.edu.hk}
\affiliation{Department of Physics, City University of Hong Kong,Tat Chee Avenue, Kowloon, Hong Kong SAR 999077, China}
\affiliation{Department of Physics, National Tsing Hua University, Hsinchu 30013, Taiwan}

\date{\today}


\begin{abstract}

Abstract: We investigate the amplification of a microwave probe signal by a superconducting artificial atom, a transmon, strongly coupled to the end of a one-dimensional semi-infinite transmission line. The end of the transmission line acts as a mirror for microwave fields. Due to the weak anharmonicity of the artificial atom, a strong pump field creates multi-photon excitations among the dressed states. Transitions between these dressed states, Rabi sidebands, give rise to either amplification or attenuation of the weak probe. We obtain a maximum amplitude amplification of about \unit[18]{\%}, higher than in any previous experiment with a single artificial atom, due to constructive interference between Rabi sidebands. We also characterize the noise properties of the system by measuring the spectrum of spontaneous emission.

\end{abstract}


\maketitle



Approaching the lowest possible noise levels, quantum-limited amplifiers play an important role in quantum information processing~\cite{bergeal2010analog, Macklin2015, grimsmo2017squeezing, kjaergaard2020superconducting} as well as other critical detection and sensing applications~\cite{danilin2018quantum, cheng2019broadband, wang2021quantum}. To better integrate such amplifiers with other circuit components, it is crucial to decrease their size, ideally down to a single atom, while maintaining desirable properties such as high gain in a broad frequency band. For an atom to fulfil these requirements, it needs to be strongly coupled to the propagating field~\cite{Tey2008, mutus2014strong}. Due to spatial mismatch in three-dimensional space, the interaction between an atom and an electromagnetic field is, in general, quite weak. However, an artificial atom confined in a one-dimensional waveguide can interact strongly with a continuum field~\cite{shen2005coherent, Astafiev2010}. A quantum amplifier based on population inversion with a single artificial atom has been realized in an open transmission line~\cite{Astafiev2010a}. Subsequently, many quantum-optical effects have been investigated in such setups, leading to the field of waveguide quantum electrodynamics (QED)~\cite{Roy2017, Sheremet2023, Hoi2011, Hoi2012, Hoi2013a, VanLoo2013, Hoi2015, Forn-Diaz2017, Wen2019, Mirhosseini2019, Wen2020, Kannan2020, lin2022deterministic}.

Existing amplification schemes in waveguide QED have employed various mechanisms, such as population inversion between bare states~\cite{Astafiev2010a}, population inversion between dressed states~\cite{Koshino2013}, and amplification without population inversion due to higher-order processes between dressed states~\cite{Wen2018}. However, none of those schemes have led to an experimentally measured amplitude amplification of more than \unit[9]{\%}. To improve on those results, we here demonstrate an increase in stimulated emission via multi-photon interference between multiple inverted excitation levels of a strongly driven superconducting artificial atom in front of a mirror. The mechanism, which builds on top of an essentially half-waveguide-QED system, differs from previous work~\cite{Koshino2013}, where the artificial atom was coupled to an open transmission line and only a two-photon pump was investigated. Due to the weak anharmonicity of the artificial atom, the strong pump field not only generates stable dressed states but also produces population inversion across them.

We investigate the cases of two-photon, three-photon, and four-photon pumping, which result in multiple Rabi sidebands that lead to either amplification or attenuation of the weak probe. When two amplification sidebands cross, the emitted photons from one sideband interfere constructively with those from the other sideband to further enhance the amplification. In addition, the half-waveguide-QED system has a unidirectional output port, which eliminates the disadvantage of losing half the stimulated emission due to the bidirectional output in a waveguide-QED system. We also engineer the artificial atom to have a relaxation rate much faster than its pure dephasing rate, which further improves the amplifying emission. With all these effects combined, we obtain a maximum amplitude amplification of about \unit[18]{\%}, which is higher than any figures reported in previous works with a single artificial atom~\cite{Astafiev2010a, Koshino2013, Wen2018}. The bandwidth and saturation power of the amplification is \unit[4]{MHz} and $\unit[-140]{dBm}$, respectively. We also characterize the noise properties of the system by analyzing the spectrum of spontaneous emission.

\begin{figure}
\includegraphics[width=\linewidth]{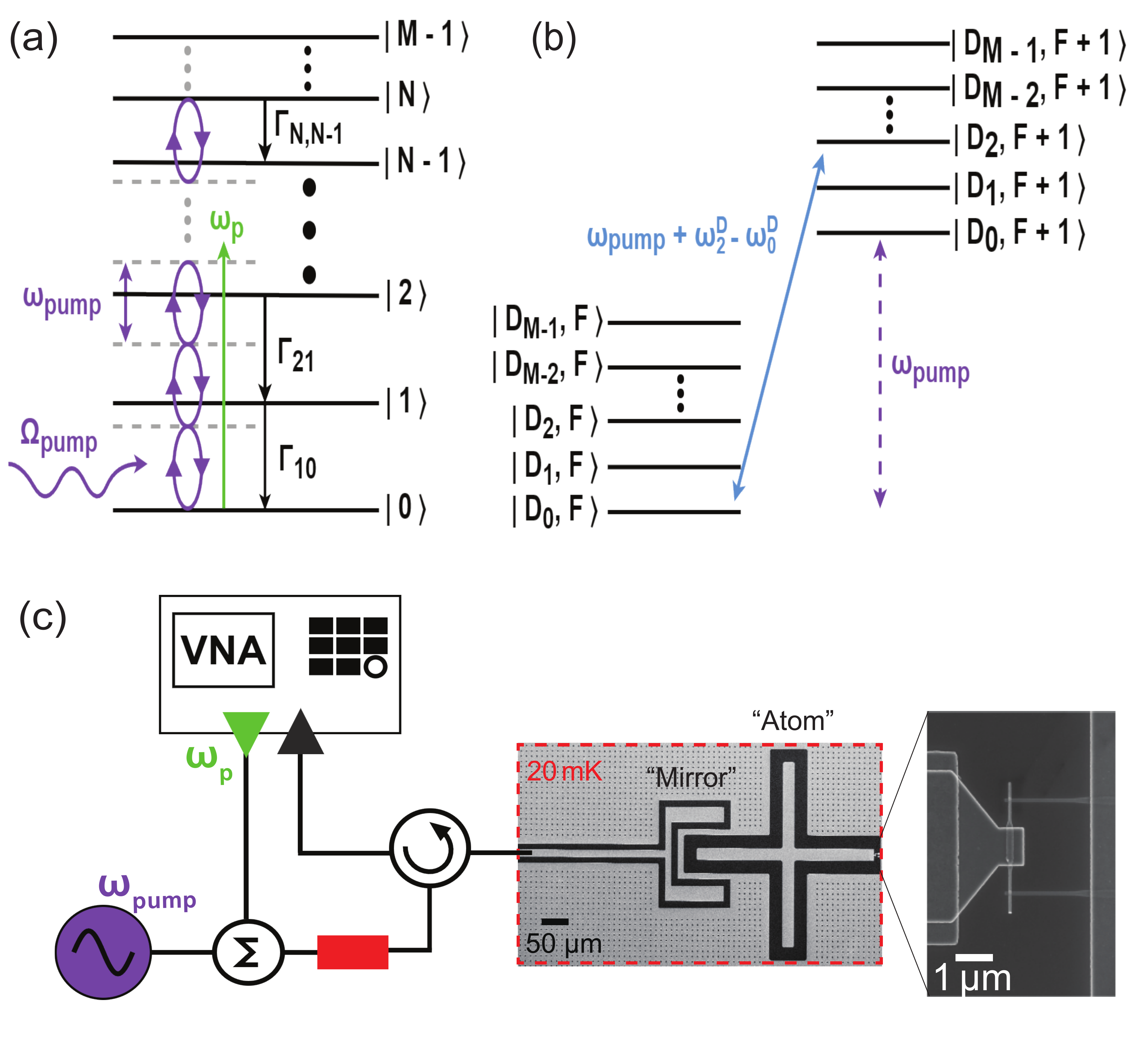}
\caption{The pump-probe scheme and the experimental setup.
(a) An $M$-level transmon is pumped by a strong resonant field with Rabi (carrier) frequency $\Omega_{\rm pump}$ ($\omega_{\rm pump}$). The transmon is pumped from $\ket{0}$ to $\ket{N}$ by an $N$-photon absorption process. A weak probe with frequency $\omega_p$ is applied to the system. The relaxation rate between adjacent states $\ket{N}$ and $\ket{N-1}$ is denoted by $\Gamma_{N, N-1}$.
(b) Energy diagram of the dressed states in the rotating frame of pump frequency $\omega_{\rm pump}$. Here $D_i$ ($i = 1, 2, \ldots, M-1$) is the $i$th eigenstate (with energy $\hbar\omega_i^D$) of the system with Hamiltonian $H_a' = H_a + H_d$ [$H_a$ and $H_d$ are defined in Eqs.~(S3) and (S4) in the Supporting Information] and $F$ denotes the photon number.
(c) A simplified circuit diagram of the experimental setup where a probe field (green) and a pump field (purple) are combined by a radio-frequency (RF) combiner at room temperature with attenuation (red rectangle) and fed into the sample (optical micrograph inside the red dashed box). The micrograph shows an artificial atom (transmon) formed by a large cross-shaped island capacitively coupled to the one-dimensional semi-infinite transmission line with a characteristic impedance of $Z_0 \simeq \unit[50]{\Omega}$. The reflected output field is measured in a vector network analyzer (VNA). The position of the superconducting quantum interference device (SQUID) loop of the transmon is shown in the scanning electron micrograph on the right. Further details on the experimental setup are given in Section S2 in the Supporting Information.
	\label{fig:multiphoton}}
\end{figure}


The experimental setup and model is illustrated in \figref{fig:multiphoton}. We consider a superconducting artificial atom, an $M$-level transmon~\cite{Koch2007}, coupled to a one-dimensional semi-infinite transmission line, with the terminated end being an anti-node mirror~\cite{Wen2018, Wen2019}. A strong resonant pump field is fed in from the open end, exciting the transmon from its ground state $\ket{0}$ to a particular excited state $\ket{N}$ by absorbing $N$ photons. In order to probe the transmon dressed by the pump, we apply a weak probe field, whose Rabi frequency $\Omega$ is much smaller than transmon decoherence rate $\gamma_{10}$, and analyze the reflection coefficient $r$ of this probe. The theoretical description of the system's dynamics and the calculation of the reflection coefficient are discussed in Section S1 in the Supporting Information.



We characterize the basic properties of the artificial atom through single-tone scattering~\cite{Hoi2015, Astafiev2010a}. We use two-tone, three-tone, and four-tone spectroscopies to characterize the energy structure of the transmon (see Section S3 in the Supporting Information for details). All the extracted parameters are summarized in \tabref{tab:Parameters}. Note that we, throughout the manuscript, calibrate the incident field by measuring the reflected field when the qubit is far detuned. In the measurements of amplification, we focus on the case of three-photon pumping, in which we see interference between Rabi sidebands. We also investigate the two- and four-photon pumping cases in Sections S4 and S5, respectively, in the Supporting Information.

\begin{table*}
	\centering
	\begin{tabular}{| c | c | c | c | c | c | c | c | c | c  | c | c |}
		\hline
		$E_C / h$  & $E_J / h$ & $E_J / E_C$ & $\omega_{10} / 2 \pi$  & $\omega_{21} / 2 \pi$ & $\omega_{32} / 2 \pi$ & $\omega_{43} / 2 \pi$  & $\Gamma_{10} / 2 \pi$  & $\Gamma_{1}^{\phi}/2\pi$ & $\gamma_{10} / 2 \pi$ \\
		\hline
		[MHz] & [GHz]  & - & [GHz] & [GHz] & [GHz] & [GHz] & [MHz]  & [MHz] & [MHz]\\
		\hline
		228 & 13.67 & 59.96 & 4.766 & 4.538 & 4.287 & 4.005 & 2.264  & 0.0317 & 1.164  \\
		\hline
	\end{tabular}
	\caption{Extracted and derived transmon parameters. We extract $\omega_{10}$, $\Gamma_{10}$, and $\gamma_{10}$ by fitting the magnitude and phase data from single-tone scattering (see Fig.~S2 in the Supporting Information). We calculate the pure dephasing rate $\Gamma_1^\phi$ from $\Gamma_{10}$ and $\gamma_{10}$, using $\gamma_{10} = \Gamma_{10} / 2 + \Gamma_1^\phi$ and the fitting technique in Ref.~\cite{lu2021characterizing}. From the four-tone spectroscopy [see Fig.~S2(c) in the Supporting Information], we extract $\omega_{21}$, $\omega_{32}$, and $\omega_{43}$, and the anharmonicity between the $\ket{0} \leftrightarrow \ket{1}$ transition and the $\ket{1} \leftrightarrow \ket{2}$ transition. The anharmonicity approximately equals the charging energy $E_C$~\cite{Koch2007}. We calculate the Josephson energy $E_J$ and $E_J / E_C$ from $\omega_{10}$ and $E_C$, where $\hbar\omega_{10} \simeq \sqrt{8 E_J E_C} - E_C$.
		\label{tab:Parameters}}
\end{table*}

\begin{figure*}[ht!]
	\includegraphics[width=\linewidth]{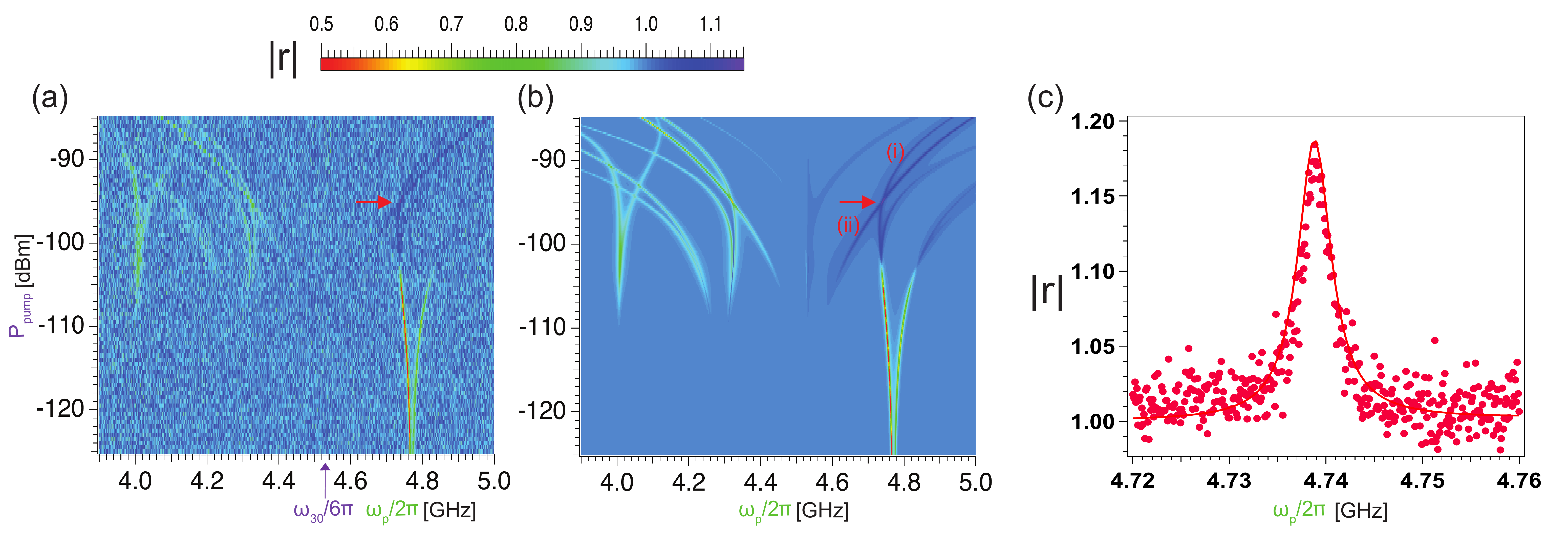}
	\caption{Reflection coefficient with three-photon pumping.
	(a) Measured magnitude $|r|$ of the reflection coefficient of the weak probe field ($P_p = \unit[-161]{dBm}$) as a function of probe frequency $\omega_p$ (x axis) and pump power $P_{\rm pump}$ (y axis). The pump frequency is $\omega_{\rm pump} = \omega_{30} / 3$.
	(b) Numerical simulation of the experiment, where we set $M = 6$ and use the relaxation rates $\Gamma_{n,n-1} / 2\pi = n \Gamma_{10} / 2\pi = \unit[2.264 n]{MHz}$ and the pure dephasing rates $\Gamma_n^\phi / 2\pi = n \Gamma_1^\phi / 2\pi = \unit[0.0317 n]{MHz}$, where $n = 1, 2, \ldots, 5$~\cite{Koch2007}. There are no free fitting parameters for the simulation.
	(c) A linecut taken at $P_{\rm pump} = \unit[-95]{dBm}$ from panels (a) and (b), where the two amplified Rabi sidebands cross each other at $\omega_p / 2\pi \approx \unit[4.739]{GHz}$ (red arrow), shows the maximum amplification to be about \unit[18]{\%}. The dots are the experimental data and the solid curve is the numerical simulation.
	\label{fig:twophoton}}
\end{figure*}


In the three-photon-pumping experiment, we pump the transition $\ket{0} \rightarrow \ket{3}$ of the transmon. When this pump is resonant, the frequencies of three pump photons sum up to $\omega_{30} = \omega_{10} + \omega_{21} + \omega_{32}$. We therefore set the pump frequency to $\omega_{\rm pump} / {2\pi} = \omega_{30} / 6\pi = \unit[4.530]{GHz}$ and sweep the pump power $P_{\rm pump}$ from $\unit[-125]{dBm}$ to $\unit[-85]{dBm}$. To probe the driven system, we simultaneously sweep a weak continuous probe field at frequency $\omega_p$ over a wide range of frequencies, including higher transitions. The experimental data for the magnitude $|r|$ of the probe reflection coefficient are shown in \figpanel{fig:twophoton}{a} and the corresponding numerical simulation are shown in \figpanel{fig:twophoton}{b}.

At low pump powers, when $P_{\rm pump} \lesssim \unit[-110]{dBm}$, only a single response is visible in the reflection spectrum, around the $\ket{0} \leftrightarrow \ket{1}$ transition frequency, in the form of split bright stripes. As the pump power increases, we observe clear and increasing splitting around the resonance frequency $\omega_{10} / 2\pi = \unit[4.766]{GHz}$. This effect is a detuned Autler-Townes splitting (detuning $\Delta = \omega_{\rm pump} - \omega_{21} = 2\pi \times \unit[8]{MHz}$)~\cite{Autler1955}, suggesting that there are two near-resonant dressed states formed by the bare states $\ket{1}$ and $\ket{2}$ with the pumping field. When the pump power increases further, beyond $\unit[-110]{dBm}$, we observe multiple Rabi sidebands, which correspond to various transitions between dressed states. The Rabi sidebands appear as amplification or attenuation of the weak probe field. In Section S6 in the Supporting Information, we analyze the Rabi sidebands for all three pumping cases and connect them to the transitions between the dressed states.


The maximum amplification occurs when the signals of the two amplified Rabi sidebands $\ket{D_3, F} \leftrightarrow \ket{D_4, F+1}$ [label (i)] and $\ket{D_4, F} \leftrightarrow \ket{D_5, F+1}$ [label (ii)] in \figpanel{fig:twophoton}{b} cross at frequency $\omega_p / 2\pi \approx \unit[4.739]{GHz}$ and pump power $P_{\rm pump} = \unit[-95]{dBm}$, interfering constructively. Figure~\figpanelNoPrefix{fig:twophoton}{c} shows a horizontal linecut at maximum amplification [indicated by red arrows in \figpanels{fig:twophoton}{a}{b}]. The amplified peak has a full width at half maximum (FWHM) of \unit[4]{MHz}. The dots (curve) are the experimental data (numerical simulation), revealing a maximum amplification of about \unit[18]{\%}. The data and the simulations are in excellent agreement without any free fitting parameters.

We note that the number of Rabi sidebands is predicted to be $N \mleft( N+1 \mright)$~\cite{Wiegand2021}. In this simplified case, where we take $N = 3$, this would yield $12$ sidebands, which is inconsistent with the measured results ($14$ sidebands). This disagreement is due to the effect of transmon higher levels than those we take into consideration.

\begin{figure*}[ht!]
	\includegraphics[width=\linewidth]{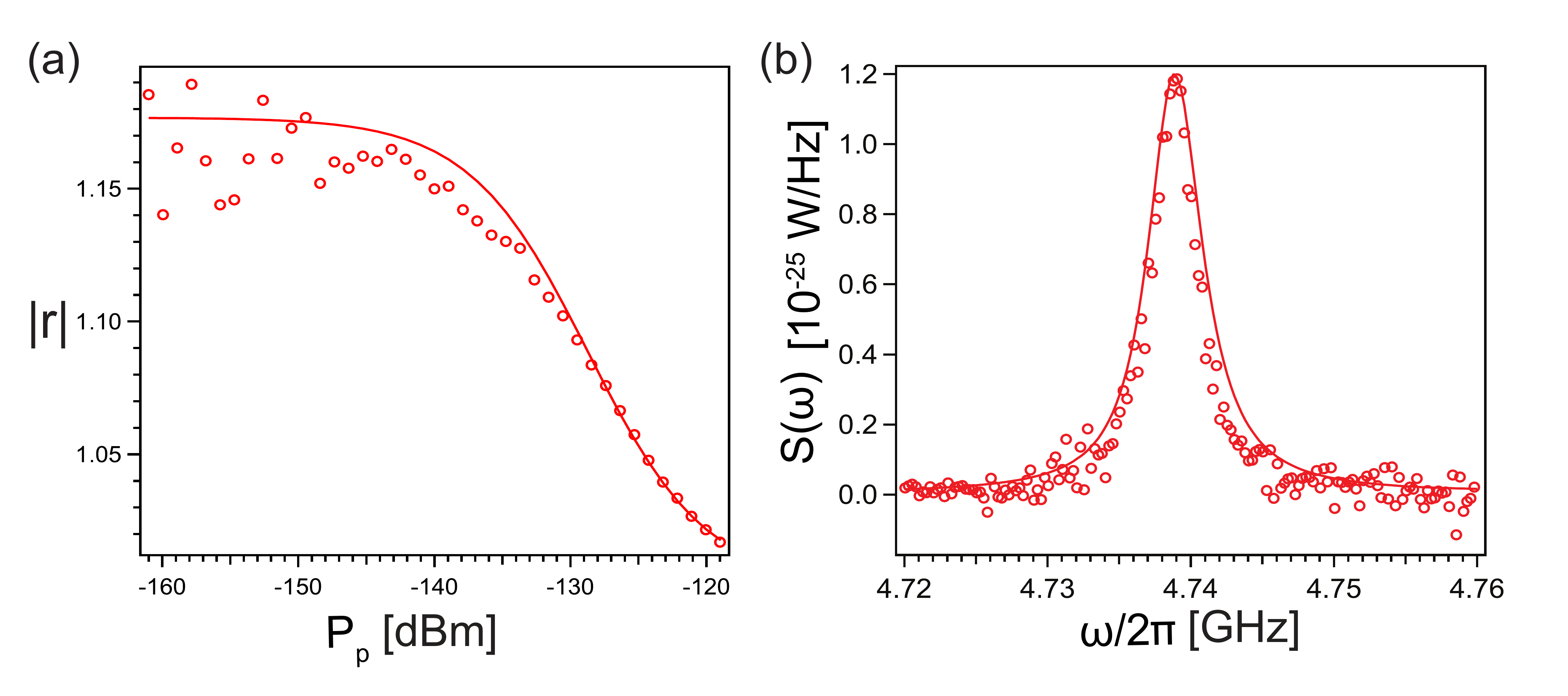}
	\caption{Saturation and noise properties of the amplification process~\cite{note2}. Open circles are experimental data; the solid red curves show the results of numerical simulations.
	(a) Magnitude $|r|$ of the reflection coefficient as a function of probe power $P_p$ at the maximum amplification point.
	(b) Spontaneous emission spectrum as a function of frequency $\omega$ at the maximum amplification point.
	\label{fig:3}}
\end{figure*}

It is important to further examine the amplification properties of the system. In \figpanel{fig:3}{a}, we show the measured saturation of the amplification process as a function of probe power $P_p$, at the maximum amplification point. The details of the measurements, including more data, are given in Section S7 in the Supporting Information. The maximum amplification of about \unit[18]{\%} is observed in the regime of weak probe power. As we increase $P_p$ beyond single-photon input, the amplification begins to saturate towards unity. To further understand the quantum noise, we also measure the spontaneous emission spectrum of the amplification using the experimental setup shown in Fig.~S1(b) in the Supporting Information. The spontaneous emission we observe in the spectral density in \figpanel{fig:3}{b} originates from the finite population of the excited states among the dressed states; it is a horizontal linecut taken from Fig.~S7(a)-(b) at $P_{\rm pump} = \unit[-96.5]{dBm}$ (details in Section S8 in the Supporting Information).



In conclusion, we investigated the amplification of a weak probe field by a single artificial atom in front of a mirror due to multi-photon excitations in dressed states induced by a strong pump field. The reflection coefficient of the weak probe was found to display multiple Rabi sidebands, where it was either amplified or attenuated. We observed a particularly strong amplification where two amplified Rabi sidebands crossed and interfered constructively. Since the atom-mirror system only has one output (instead of two like an open waveguide) and we employed a high-coherence transmon, where the relaxation rate to the waveguide was significantly higher than the pure dephasing rate, the amplification was further enchanced. As a result, we observed an amplification of about \unit[18]{\%}, which is almost half of the fundamental limit of stimulated emission ($\sqrt{2}$). The bandwidth and saturation power of the amplification were \unit[4]{MHz} and $\unit[-140]{dBm}$, respectively. Our results demonstrate potential for the development of integrated and scalable quantum amplifiers and have implications for the advancement of quantum technologies such as sensing, computation, and communication.


\section{Supporting Information}

The Supporting Information is available free of charge on the ACS Publications website. It contains the details about the following subjects: system dynamics and calculation on reflection coefficient, experimental setups, spectroscopy for a transmon in front of a mirror, the case of two-photon pumping, the case of four-photon pumping, Rabi sidebands for multi-photon pumping, saturation power of amplification, and noise properties of amplification.


\section{Acknowledgements}

I.-C.H.~acknowledges financial support from City University of Hong Kong through the start-up project 9610569, and from the Research Grants Council of Hong Kong (Grant number 11312322).
K.-T.L. and G.D.L.~acknowledges support from NSTC of Taiwan under projects 110-2112-M-002 -026, 111-2112-M-002-037, and 111-2811-M-002-087, and NTU under project NTU-CC-110L890106.
A.F.K. acknowledges support from the Swedish Research Council (grant number 2019-03696), from the Swedish Foundation for Strategic Research, and from the Knut and Alice Wallenberg Foundation through the Wallenberg Centre for Quantum Technology (WACQT).
J.C.C.~acknowledges financial support from the NSTC of Taiwan under projects 110-2112-M-007-022-MY3 and 111-2119-M-007-008.
H.I.~acknowledges support from FDCT Macau under grants 0130/2019/A3 and 0015/2021/AGJ and support from University of Macau under grant MYRG2018-00088-IAPME.
P.Y.W.~acknowledges financial support from the NSTC of Taiwan under project 110-2112-M-194-006-MY3.


\normalem
\bibliography{Ref}

\begin{thebibliography}{32}
\bibitem [{not()}]{note2}%
  \BibitemOpen
  \href@noop {} {}\bibinfo {note} {Figure 2(a) and Figure 3 were measured in
  different cool-downs with the same sample.}\BibitemShut {Stop}%
\end{thebibliography}
\begin{figure}[ht!]
\includegraphics[width=\linewidth]{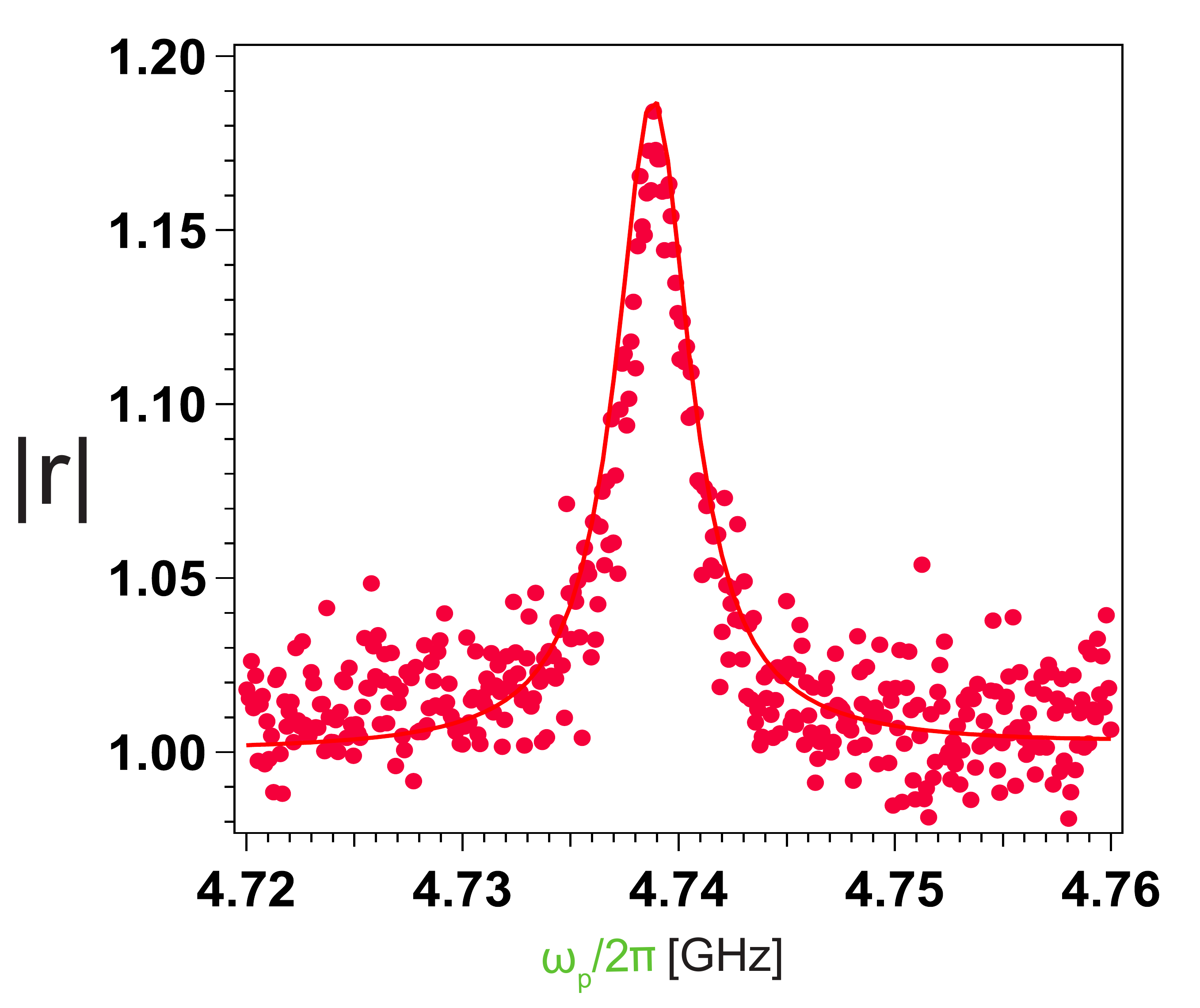}
\caption{TOC graphic.
\label{fig:TOC}}
\end{figure}

\end{document}


\title{Supporting Information for ``Microwave amplification via interfering multi-photon processes in a half-waveguide quantum electrodynamics system''}

\author{Fahad Aziz}
\thanks{F.~A, K.~T.~L, and P.~Y.~W contributed equally to this work.}
\affiliation{Department of Physics, National Tsing Hua University, Hsinchu 30013, Taiwan}

\author{Kuan Ting Lin}
\thanks{F.~A, K.~T.~L, and P.~Y.~W contributed equally to this work.}
\affiliation{Department of Physics and CQSE, National Taiwan University, Taipei 10617, Taiwan}

\author{Ping Yi Wen}
\thanks{F.~A, K.~T.~L, and P.~Y.~W contributed equally to this work.}
\affiliation{Department of Physics, National Chung Cheng University, Chiayi 621301, Taiwan}

\author{Samina}
\affiliation{Department of Physics, National Tsing Hua University, Hsinchu 30013, Taiwan}

\author{Yu Chen Lin}
\affiliation{Department of Physics and CQSE, National Taiwan University, Taipei 10617, Taiwan}

\author{Emely Wiegand}
\affiliation{Department of Microtechnology and Nanoscience, Chalmers University of Technology, 412 96 Gothenburg, Sweden}

\author{Ching-Ping Lee}
\affiliation{Department of Physics, National Tsing Hua University, Hsinchu 30013, Taiwan}

\author{Yu-Ting Cheng}
\affiliation{Department of Physics, National Tsing Hua University, Hsinchu 30013, Taiwan}

\author{Ching-Yeh Chen}
\affiliation{Department of Physics, National Tsing Hua University, Hsinchu 30013, Taiwan}

\author{Chin-Hsun Chien}
\affiliation{Department of Physics, National Tsing Hua University, Hsinchu 30013, Taiwan}

\author{Kai-Min Hsieh}
\affiliation{Department of Physics, National Tsing Hua University, Hsinchu 30013, Taiwan}

\author{Yu-Huan Huang}
\affiliation{Department of Physics, National Tsing Hua University, Hsinchu 30013, Taiwan}

\author{Ian Hou}
\affiliation{Institute of Applied Physics and Materials Engineering, University of Macau, Macau}
\affiliation{UMacau Zhuhai Research Institute, Zhuhai, Guangdong, China}

\author{Jeng-Chung Chen}
\affiliation{Department of Physics, National Tsing Hua University, Hsinchu 30013, Taiwan}
\affiliation{Center for Quantum Technology, National Tsing Hua University, Hsinchu 30013, Taiwan}

\author{Yen-Hsiang Lin}
\affiliation{Department of Physics, National Tsing Hua University, Hsinchu 30013, Taiwan}
\affiliation{Center for Quantum Technology, National Tsing Hua University, Hsinchu 30013, Taiwan}

\author{Anton Frisk Kockum}
\affiliation{Department of Microtechnology and Nanoscience, Chalmers University of Technology, 412 96 Gothenburg, Sweden}

\author{Guin Dar Lin}
\affiliation{Department of Physics and CQSE, National Taiwan University, Taipei 10617, Taiwan}
\affiliation{Physics Division, National Center for Theoretical Sciences, Taipei 10617, Taiwan}
\affiliation{Trapped-Ion Quantum Computing Laboratory, Hon Hai Research Institute, Taipei 11492, Taiwan}

\author{Io-Chun Hoi}
\email{iochoi@cityu.edu.hk}
\affiliation{Department of Physics, City University of Hong Kong,Tat Chee Avenue, Kowloon, Hong Kong SAR 999077, China}
\affiliation{Department of Physics, National Tsing Hua University, Hsinchu 30013, Taiwan}

\date{\today}

\maketitle

\tableofcontents

\renewcommand{\thefigure}{S\arabic{figure}}
\renewcommand{\thesection}{S\arabic{section}}
\renewcommand{\theequation}{S\arabic{equation}}
\renewcommand{\thetable}{S\arabic{table}}
\renewcommand{\bibnumfmt}[1]{[S#1]}
\renewcommand{\citenumfont}[1]{S#1}


\newpage

\section{System dynamics and calculation on reflection coefficient}
\label{sec:theory}

The dynamics of such a system is described by the quantum master equation,
\begin{align}
	\frac{d\rho}{dt}= & -\frac{i}{\hbar}\left[H_{S},\rho\right]\label{eq: master_equation-1}\\
	& +\sum_{n,m=1}^{M-1}\frac{\Gamma_{n,n-1}+\Gamma_{m,m-1}}{2}\mathcal{D}\left[\sigma_{n,n-1},\sigma_{m-1,m}\right]\rho+\sum_{n=1}^{M-1}2\Gamma_{n}^{\phi}\mathcal{D}\left[\sigma_{n,n},\sigma_{n,n}\right]\rho,\nonumber 
\end{align}
where the system Hamiltonian is denoted by 
\begin{equation}
	H_{S}=H_{a}+H_{d}+H_{p},\label{eq: H_s-1}
\end{equation}
with
\begin{align}
	H_{a} & =\sum_{n=1}^{M-1}\hbar\left(\omega_{n}-n\omega_\text{pump}\right)\sigma_{n,n},\label{eq: H_a-1}\\
	H_{d} & =\sum_{n=1}^{M-1}\sqrt{n}\frac{\hbar\Omega_\text{pump}}{2}\sigma_{n,n-1}+\text{H.c.},\label{eq: H_d-1}\\
	H_{p} & =\sum_{n=1}^{M-1}\sqrt{n}\frac{\hbar\Omega_{p}}{2}\sigma_{n,n-1}e^{-i\left(\omega_{p}-\omega_\text{pump}\right)t}+\text{H.c.}.\label{eq: H_p-1}
\end{align}
Here, $\sigma_{n,n}=\left|n\right\rangle \left\langle n\right|$ is
the projection operator for the $n$th energy level with atomic energy
$\hbar\omega_{n}$ and $\sigma_{n,n-1}=\left|n\right\rangle \left\langle n-1\right|$
is the atomic ladder operator between the $n$th and $\left(n-1\right)$th
level of the transmon. The Rabi frequency and the carrier frequency
of the pump (probe) field are denoted by $\Omega_\text{pump}$ $\left(\Omega_{p}\right)$
and $\omega_\text{pump}$ $\left(\omega_{p}\right)$, respectively. The Lindblad
superoperator is denoted by $\mathcal{D}\left[A,B\right]\rho=B\rho A-\frac{1}{2}AB\rho-\frac{1}{2}\rho AB$.
The relaxation rate between the levels $\left|n\right\rangle $ and
$\left|n-1\right\rangle $ is given by $\Gamma_{n,n-1}$. The last
term in Eq. $\left(\ref{eq: master_equation-1}\right)$ is added to
account for the pure dephasing process with the dephasing rate, $\Gamma_{n}^{\phi}$,
for the $n$th level and $\text{H.c.}$ stands for Hermitian conjugate.

The reflection coefficient is determined by $r = \mleft| \frac{\expec{a_{\rm out}}}{\expec{a_{\rm in}}}\mright|$, where the output and input signals are the output annihilation operator $a_{\rm out}$ and the input one $a_{\rm in}$, respectively~\cite{Gardiner1985}. The output signal can be determined from the input signal and the atomic response via the input-output relation~\cite{Lalumiere2013}
\begin{equation}
	a_{\rm out} (t) = a_{\rm in} (t) - \sum_{n=1}^{M-1} \sqrt{\Gamma_{n,n-1}} \sigma_{n-1,n} (t)
	\label{eq:input_outout_relation}.
\end{equation}
In our setup, we apply a single-mode classical probe field as an input signal. Then, the input operator can be approximated by a classical field~\cite{strandberg2019numerical, lu2021propagating}
\begin{equation}
	a_{\rm in} (t) \rightarrow - \frac{i \Omega_p}{2 \sqrt{\Gamma_{10}}} e^{-i \mleft( \omega_p - \omega_\text{pump} \mright) t}\label{eq:classical probe beam}.
\end{equation}
Hence, the reflection coefficient due to the weak probe beam is given by
\begin{equation}
	r = \mleft| 1 - i \sum_{n=1}^{M-1} \frac{\sqrt{\Gamma_{10} \Gamma_{n,n-1}} \expec{\sigma_{n-1,n} (t)}}{\Omega_p} \mright|\label{eq: reflection}.
\end{equation}
%


\newpage

\section{Experimental setups}

\begin{figure}[h!]
	\includegraphics[width=\linewidth]{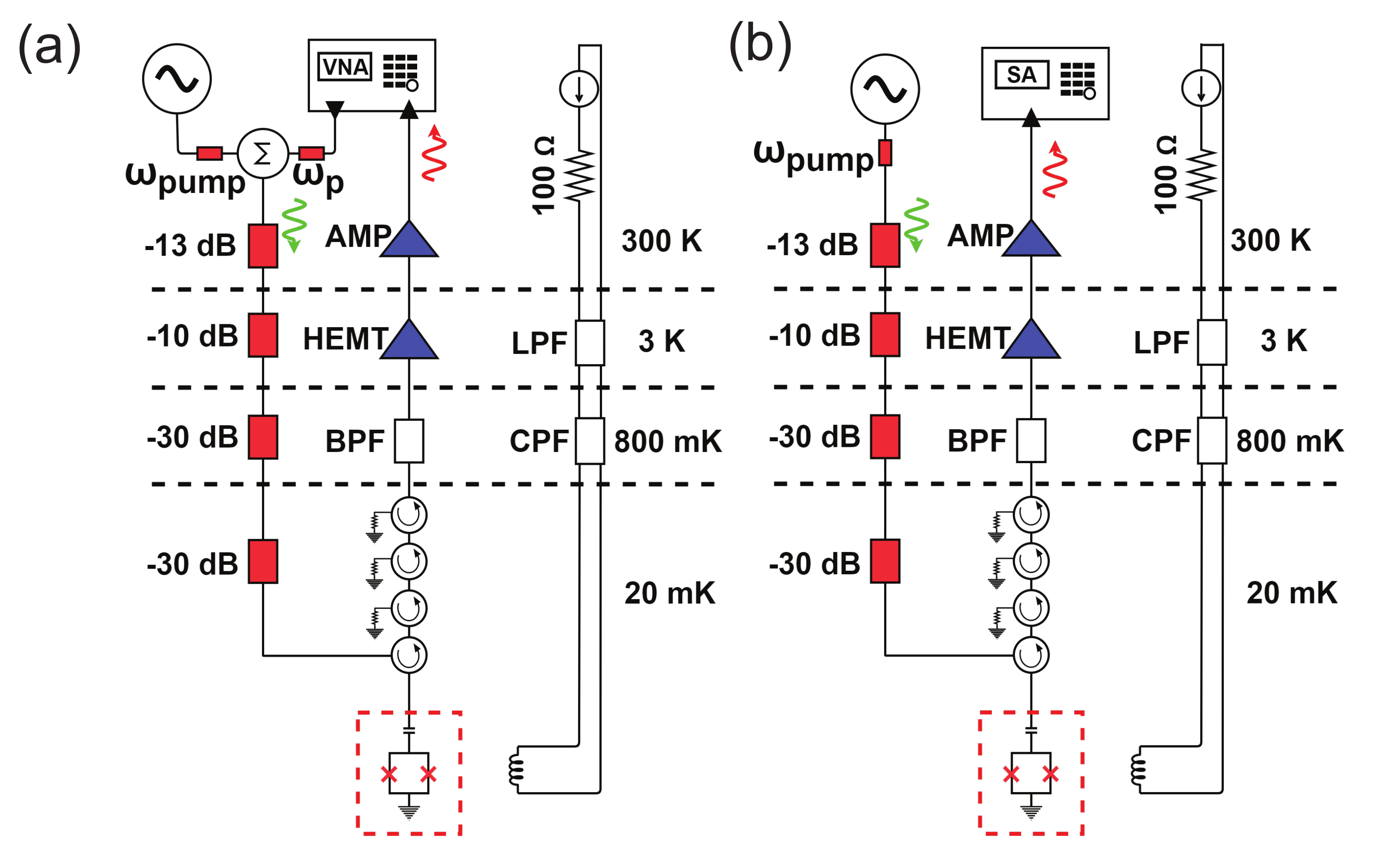}
	\caption{Schematics of the experimental setups with an artificial atom in a one-dimensional semi-infinite transmission line for (a) measuring the reflection coefficient $r$ and (b) measuring the atomic emission spectrum. In the setup in (a), a weak probe microwave field at frequency $\omega_p$ from the vector network analyzer (VNA) and a strong pump field at frequency $\omega_{\rm pump}$ from the radio-frequency (RF) source are applied through a RF combiner ($\Sigma$) at room temperature and fed into the input through several attenuators (solid red rectangles) at each stage of the dilution refrigerator, and a cryogenic circulator. The output reflected signal from the sample is passed through several cryogenic circulators in series, a band-pass filter (BPF), a high-electron-mobility transistor (HEMT) amplifier located in the dilution refrigerator (\unit[3]{K}), and further amplified at room temperature, then measured with the VNA. A small superconducting magnet attached to the sample box is used to flux bias the artificial atom, controlled by a direct-current (DC) source at room temperature. In the setup in (b), a spectrum analyzer (SA) measures the spectrum of emission from the artificial atom by pumping it strongly at frequency $\omega_{\rm pump}$ from an RF source.
	\label{fig:Device}}
\end{figure}

The schematics of the experimental setups are shown in \figref{fig:Device}. Our device (inside the dashed red rectangle) consists of a superconducting artificial atom (transmon), strongly coupled to a one-dimensional semi-infinite transmission line with characteristic impedance $Z_0 \simeq \unit[50]{\Omega}$. This is equivalent to putting an atom in front of a mirror~\cite{Hoi2015}. The transition frequency between the energy levels $\ket{0}$ and $\ket{1}$ of a transmon qubit is given by $\hbar \omega_{10}(\Phi)\simeq \sqrt{8 E_C E_J(\Phi)} - E_C$, which is determined by the charging energy $E_C = e^2 / 2C_\Sigma$, where $C_\Sigma$ is the total capacitance and $e$ is the elementary charge, and the Josephson energy $E_J (\Phi)$, where $\Phi$ is the external magnetic flux.

\newpage

\section{Spectroscopy for a transmon in front of a mirror}

\begin{figure}[h!]
	\includegraphics[width=\linewidth]{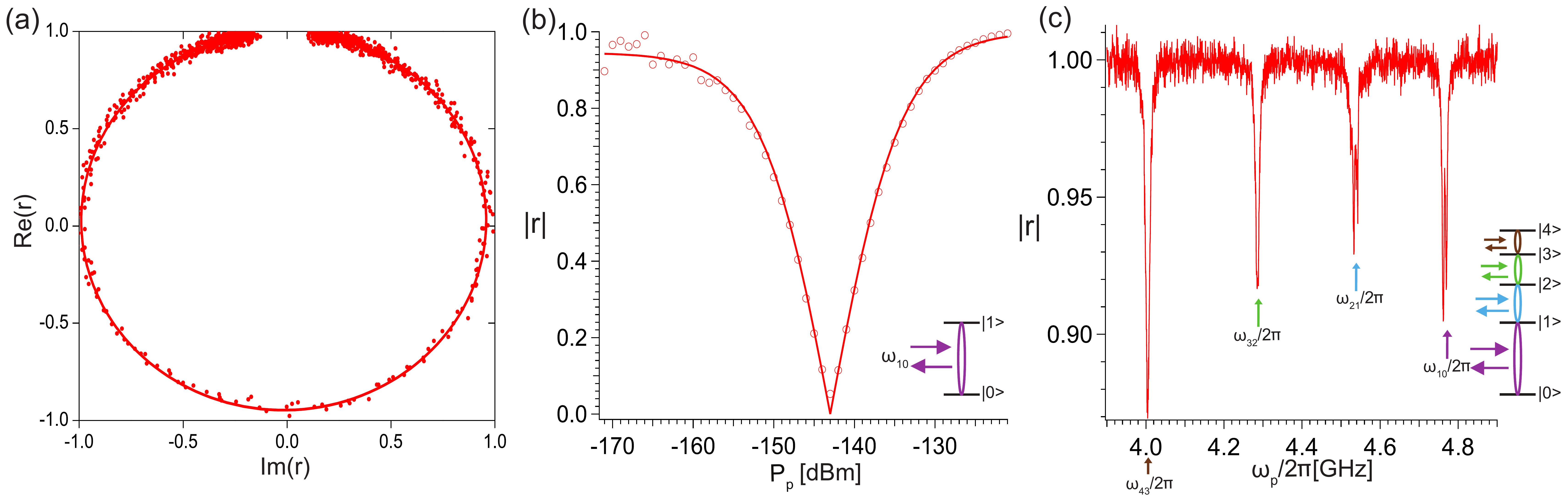}
	\caption{Single- and multi-tone spectroscopy.
	(a) The measured reflection coefficient $r$ plotted in a complex-plane representation at weak probe power $P_p = \unit[-161]{dBm}$. The experimental data (dots) are fit (solid curve) simultaneously by using the circle-fit technique~\cite{Probst2015, lu2021characterizing}.
	(b) The magnitude $|r|$ of the reflection coefficient as a function of incident resonant probe power (red open circles for experimental data; red solid curve for the fit). The inset shows the on-resonant incident and reflected fields of the two-level atom.
	(c) Four-tone spectroscopy of the transmon for revealing higher transitions ($\omega_{21}$, $\omega_{32}$, and $\omega_{43}$). The vertical arrows of different colors below each dip in the reflection coefficient $|r|$ correspond to the resonant transition frequency in the energy structure of a multi-level atom (inset).
	\label{fig:spectroscopy}}
\end{figure}

We perform single-tone spectroscopy to characterize the basic parameters of the artificial atom. We measure the reflection coefficient $r$ as a function of probe frequency $\omega_p$ and probe power $P_p$. With a weak probe, where the Rabi frequency is much smaller than the decoherence rate ($\Omega \ll \gamma_{10}$), the reflection coefficient $r$ in the complex plane resembles a perfect circle. In \figpanel{fig:spectroscopy}{a}, we extract the resonance frequency $\omega_{10}$, the relaxation rate $\Gamma_{10}$, and the decoherence rate $\gamma_{10}$. In \figpanel{fig:spectroscopy}{b}, by increasing the power of the resonant microwave, we see the non-linear scattering properties of the two-level atom. At the point where $|r| = 0$, the power is $\hbar \omega_{10} \Gamma_{10} / 8$. The parameters are extracted using the circle-fit technique in Refs.~\cite{Probst2015,lu2021characterizing} and summarized in Table I in the main text.

Next, to characterize the energy structure of our artificial atom, we bias it at a fixed transition frequency ($\omega_{10} / 2\pi = \unit[4.766]{GHz}$) by using a superconducting magnet (flux). To perform two-tone spectroscopy, we superimpose the pump field at frequency $\omega_{\rm pump}$ from a radio-frequency (RF) source with the probe field at frequency $\omega_p$ from the vector network analyzer (VNA) using an RF combiner. Specifically, we apply a strong constant microwave pump at frequency $\omega_{\rm pump} = \omega_{10}$ to saturate the $\ket{0} \rightarrow \ket{1}$ transition and measure the reflection of a weak probe, sweeping the probe frequency $\omega_p$. The population of the first excited state increases, resulting in photon scattering from the $\ket{1}$$\leftrightarrow$$\ket{2}$ transition, showing as scattering  $\omega_{\rm p}/{2\pi}=\omega_{21}/{2\pi}=\unit[4.538]{GHz}$. From this, we determine the qubit anharmonicity to be  $\alpha/{2\pi} \equiv \mleft(\omega_{21} - \omega_{10} \mright) / 2\pi = \unit[-228]{MHz}$. To characterize the $\ket{2} \rightarrow \ket{3}$ transition frequency, we perform three-tone spectroscopy, adding another strong constant microwave pump at frequency $\omega_{\rm pump2}$ = $\omega_{21}$. The measured transition frequency is $\omega_{32} / 2\pi = \unit[4.287]{GHz}$. To probe all the $\ket{0} \leftrightarrow \ket{1}$, $\ket{1} \leftrightarrow \ket{2}$, $\ket{2} \leftrightarrow \ket{3}$, and $\ket{3} \leftrightarrow \ket{4}$ transitions, we performed four-tone spectroscopy, shown in \figpanel{fig:spectroscopy}{c}. Here, we apply three calibrated strong constant microwave pumps at frequencies $\omega_{\rm pump1}$ = $\omega_{10}$, $\omega_{\rm pump2}$ = $\omega_{21}$, and $\omega_{\rm pump3}$ = $\omega_{32}$. The extracted parameters are summarized in Table I in the main text.

\newpage

\section{Two-photon pumping case}

\begin{figure}[h!]
	\includegraphics[width=\linewidth]{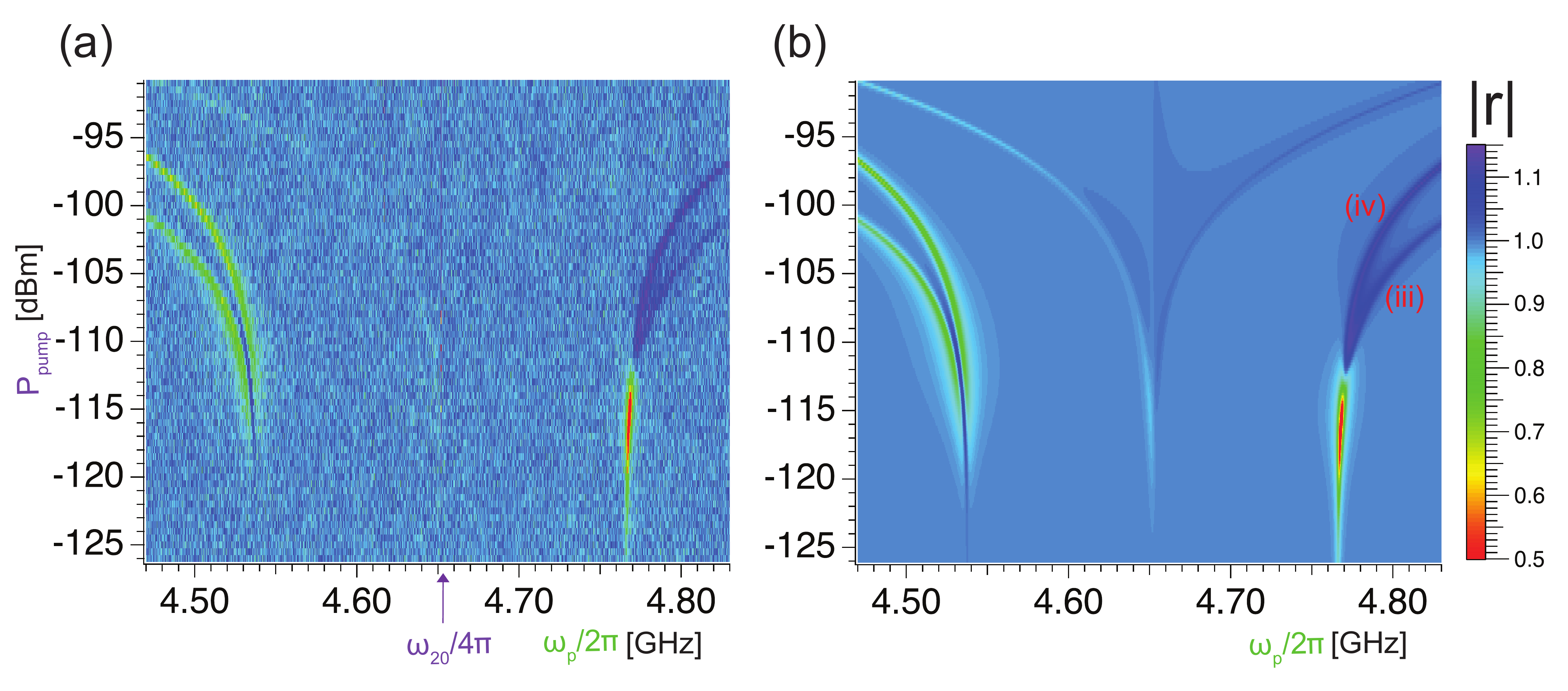}
	\caption{The magnitude $|r|$ of the reflection coefficient for a weak probe with a two-photon pump signal. (a) Experimental and (b) numerical plots of $|r|$ as a function of probe frequency $\omega_p$ (x axis) and pump power $P_{\rm pump}$ (y axis). We apply a fixed microwave pump at $\omega_{\rm pump} = \omega_{20}/{2}$. With increasing pump power, the bare-state transitions split into the dressed-state transitions, which appear symmetrically with respect to the pump frequency as amplified or attenuated Rabi sidebands of the probe field.
	\label{fig:7}}
\end{figure}

The experiment whose results are shown in Figure~\figpanelNoPrefix{fig:7}{a} is performed on the three lowest energy levels ($\ket{0}$, $\ket{1}$, and $\ket{2}$) of a ladder-type three-level superconducting artificial atom, with transition frequencies $\omega_{10}$ and $\omega_{12}$. We pump our system with a two-photon process at frequency $\omega_{\rm pump}/{2\pi}$ = $\omega_{20}/{4\pi}=\unit[4.652]{GHz}$ and vary the pump power $P_\text{pump}$ (y axis), along with sweeping a continuous weak probe field at frequency $\omega_p$ (x axis). As a result, six Rabi sidebands appear, associated with dressed states under a strong pump power~\cite{Koshino2013}. The reflected signals of the Rabi sidebands below the pump frequency regime are attenuated ($|r| < 1$), while the Rabi sidebands above the pump frequency regime are amplified ($|r| > 1$). This amplification is the result of the population inversion that occurs within the dressed states of the strongly pumped three-level system. The most significant amplification, about \unit[15]{\%}, is seen when the probe is close to resonant with one of the dressed-state transitions at $P_{\rm pump} = \unit[-103]{dBm}$. This amplification is about twice that measured in the work using an open transmission line~\cite{Koshino2013}. This increase is because of the mirror, which enhances amplification primarily by directing all atomic output in one direction.

The corresponding numerical result is shown in \figpanel{fig:7}{b}. See \secref{sec:theory} for details of the simulations. At pump powers $P_{\rm pump} < \unit[-120]{dBm}$, the pump field is insufficient to significantly excite the transmon from the ground state to higher ones, meaning that one can only observe the $\ket{0} \leftrightarrow \ket{1}$ transition near the frequency $\omega_p / 2\pi \approx \unit[4.766]{GHz}$. Conversely, for $P_{\rm pump} > \unit[-120]{dBm}$, the higher levels start to be populated, implying that the $\ket{1} \leftrightarrow \ket{2}$ transition can be observed at $\omega_p / 2\pi \approx \unit[4.54]{GHz}$.
In addition, one can observe six Rabi sideband signals when $P_{\rm pump} = \unit[-103]{dBm}$, originating from the transitions between dressed states~\cite{Koshino2013}, as shown in Figure 1(b) in the main text.
In that case, it is worth noting that the reflected signals from the $\ket{D_3,F} \leftrightarrow \ket{D_5,F+1}$ [label (iii)] and $\ket{D_4,F} \leftrightarrow \ket{D_5,F+1}$ [label (iv)] dressed-state transitions (Rabi sidebands) are amplified by about \unit[15]{\%} due to the dressed-state population inversion~\cite{Wiegand2021}. There is a good agreement between the experimental data and the numerical simulation of the reflection coefficient without any free fitting parameter.

\newpage

\section{Four-photon pumping case}

\begin{figure}[h!]
	\includegraphics[width=\linewidth]{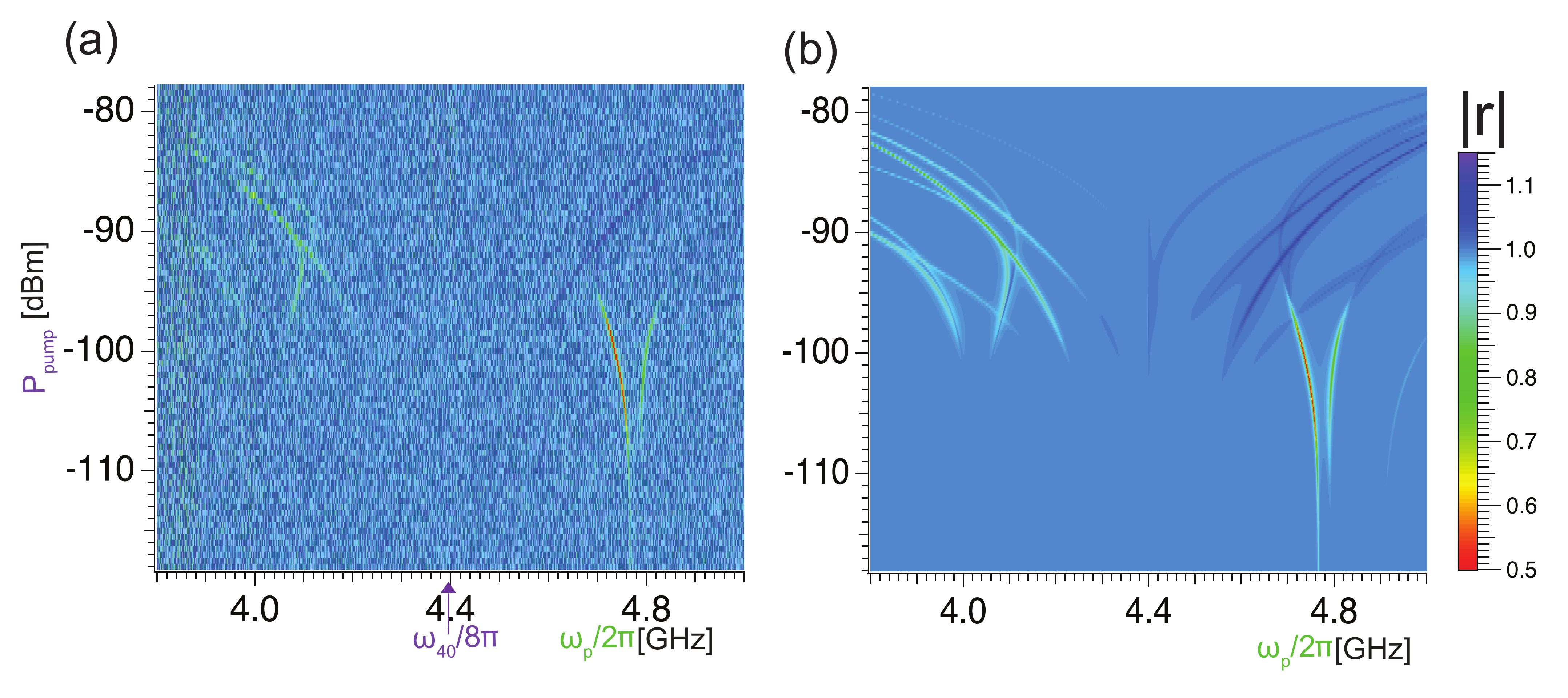}
	\caption{The magnitude $|r|$ of the reflection coefficient for a weak probe with a four-photon pump signal. (a) Experimental and (b) numerical plots of $|r|$ as a function of probe frequency $\omega_p$ (x axis) and pump power $P_{\rm pump}$ (y axis). We apply a fixed microwave pump at $\omega_{\rm pump} = \omega_{40} / 4$. At low pump power, we observe only two Rabi sidebands around frequency $\omega_{10}$, and at a higher pump power regime, as expected, we find multiple Rabi sidebands from higher transitions, which reflect amplified or attenuated Rabi sidebands of the weak probe field.
	\label{fig:8}}
\end{figure}

In \figpanel{fig:8}{a} [\figpanelNoPrefix{fig:8}{b}], we plot the magnitude of the reflection coefficient measured in a four-photon pump experiment [numerical simulation]. We pump our system with a four-photon process at frequency $\omega_{\rm pump} / {2\pi} = \omega_{40} / 8\pi \approx \unit[4.399]{GHz}$ and vary the pump power $P_{\rm pump}$ along with sweeping a continuous weak probe field at frequency $\omega_p$ from \unit[3.8]{GHz} to \unit[5]{GHz}. When $P_{\rm pump} \lesssim \unit[-100]{dBm}$, we observe only two Rabi sidebands. Conversely, when $P_{\rm pump} > \unit[-100]{dBm}$, the higher energy levels start to be populated, resulting in the observation of multiple Rabi sidebands as amplified or attenuated sidebands. We observe a couple of Rabi sidebands that are quite faint in the experimental data, but which are clearly visible in the numerical simulation. The limited visibility in the experimental data is due to noise.

\newpage

\section{Rabi sidebands for multi-photon pumping}

\begin{figure}[h!]
	\includegraphics[width=\linewidth]{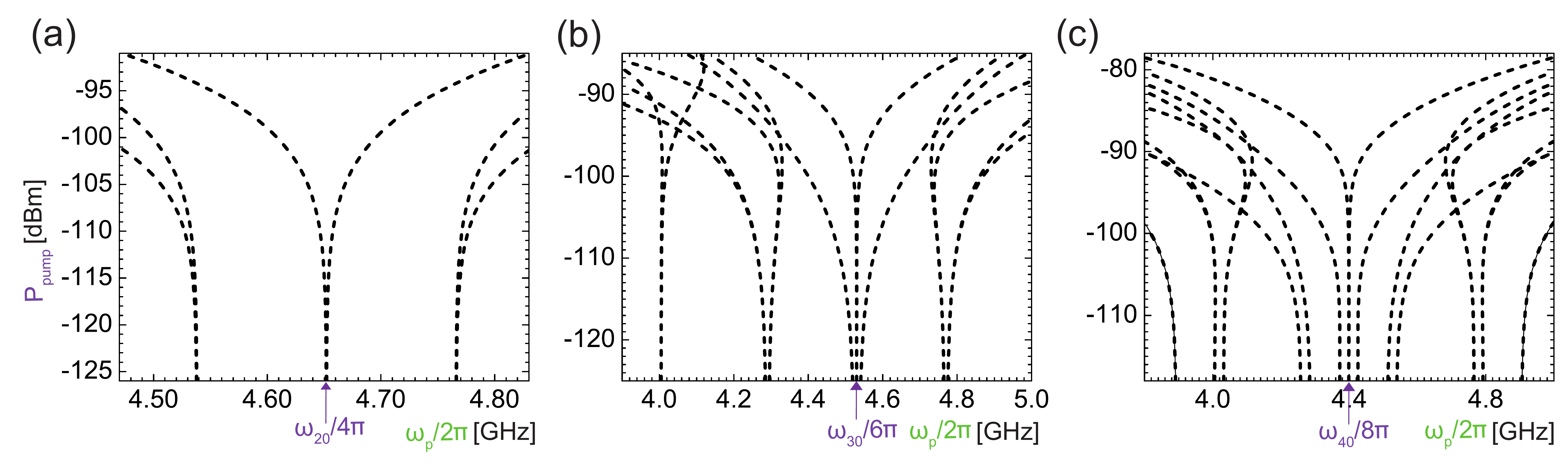}
	\caption{Numerically calculated Rabi sidebands due to multi-photon pumping, as a function of the probe frequency $\omega_p$ and the pump power $P_{\rm pump}$. (a) Two-photon pumping, (b) three-photon pumping, and (c) four-photon pumping. The vertical arrow in each plot shows the applied constant microwave pump at $\omega_{\rm pump}$.
	\label{fig:rabibands}}
\end{figure}

Our numerical analysis of all Rabi sidebands for two-photon, three-photon, and four-photon pumping is conducted using the parameters described in the main text. The resulting predicted sidebands are shown in \figref{fig:rabibands}. The transitions between the dressed states for two-photon, three-photon, and four-photon pumping are shown in Table~\ref{tab:2}. The ordering of the transitions in the table corresponds to the dashed curves from right to left in each panel of \figref{fig:rabibands}. The Rabi sidebands for two-photon pumping are shown in \figpanel{fig:rabibands}{a} as a function of the probe frequency $\omega_p$ and the pump power $P_{\rm pump}$. At $P_{\rm pump} = \unit[-115]{dBm}$, the three transitions (i.e., from right to left, $\omega_{10}$, $\omega_{20}/2$, and $\omega_{21}$) begin to split into the dressed states, resulting in six Rabi sidebands.
Likewise, \figpanel{fig:rabibands}{b} shows all the Rabi sidebands for three-photon pumping; when $P_{\rm pump} > \unit[-100]{dBm}$, 14 Rabi sidebands are observed.
Finally, \figpanel{fig:rabibands}{c} shows the Rabi sidebands for four-photon pumping; when $P_{\rm pump} > \unit[-95]{dBm}$, we observe 20 Rabi sidebands appearing symmetrically around the pump frequency $\omega_{\rm pump} / {2\pi} = \omega_{40} / 8\pi$. Note that the highest and lowest Rabi sideband have two transitions that overlap each other.

\newpage

\begin{table}	
	\small\addtolength{\tabcolsep}{50pt}
	\begin{tabular}{  l l   } 
		\hline \hline
		Pumping cases	    &   Transitions between dressed states   (label)   \\
		\hline 
		1. Two-photon pumping & 1. \textcolor{blue}{ $\left|D_{3},F\right\rangle \longleftrightarrow\left|D_{5},F+1\right\rangle $   \hspace*{1.5cm}}(iii)  \\ & 2. \textcolor{blue} { $\left|D_{4},F\right\rangle \longleftrightarrow\left|D_{5},F+1\right\rangle $ \hspace*{1.5cm}}(iv) \\ & 3. \textcolor{blue}{ $\left|D_{3},F\right\rangle \longleftrightarrow\left|D_{4},F+1\right\rangle $} \\ & 4. \textcolor{red}{ $\left|D_{4},F\right\rangle \longleftrightarrow\left|D_{3},F+1\right\rangle $} \\ & 5. \textcolor{red}{  $\left|D_{5},F\right\rangle \longleftrightarrow\left|D_{4},F+1\right\rangle $} \\ & 6. \textcolor{red}{ $\left|D_{5},F\right\rangle \longleftrightarrow\left|D_{3},F+1\right\rangle $} \\
		\hline
		 2. Three-photon pumping & 1. \textcolor{blue}{ 
		$\left|D_{2},F\right\rangle \longleftrightarrow\left|D_{5},F+1\right\rangle $} \\ & 2. \textcolor{blue}{ $\left|D_{3},F\right\rangle \longleftrightarrow\left|D_{5},F+1\right\rangle $} \\ & 3. \textcolor{blue}{ $\left|D_{2},F\right\rangle \longleftrightarrow\left|D_{4},F+1\right\rangle $} \\ & 4. \textcolor{blue}{ $\left|D_{3},F\right\rangle \longleftrightarrow\left|D_{4},F+1\right\rangle $}  \hspace*{1.5cm}(i)\\ & 5. \textcolor{blue}{ $\left|D_{4},F\right\rangle \longleftrightarrow\left|D_{5},F+1\right\rangle $} \hspace*{1.5cm}(ii)\\ & 6. \textcolor{blue}{ $\left|D_{2},F\right\rangle \longleftrightarrow\left|D_{3},F+1\right\rangle $} \\  & 7. \textcolor{red}{ $\left|D_{3},F\right\rangle \longleftrightarrow\left|D_{2},F+1\right\rangle $} \\ & 8. \textcolor{red}{ $\left|D_{5},F\right\rangle \longleftrightarrow\left|D_{4},F+1\right\rangle $} \\ & 9. \textcolor{red}{ $\left|D_{4},F\right\rangle \longleftrightarrow\left|D_{3},F+1\right\rangle $} \\ & 10. \textcolor{red}{ $\left|D_{4},F\right\rangle \longleftrightarrow\left|D_{2},F+1\right\rangle $} \\ & 11. \textcolor{red}{ $\left|D_{5},F\right\rangle \longleftrightarrow\left|D_{3},F+1\right\rangle $} \\ & 12. \textcolor{red}{ $\left|D_{5},F\right\rangle \longleftrightarrow\left|D_{2},F+1\right\rangle $} \\ & 13. \textcolor{red}{ $\left|D_{2},F\right\rangle \longleftrightarrow\left|D_{1},F+1\right\rangle $} \\ & 14. \textcolor{red}{ $\left|D_{3},F\right\rangle \longleftrightarrow\left|D_{1},F+1\right\rangle $ }\\ 
		
		\hline
		 3. Four-photon pumping & 1. \textcolor{red}{ $\left|D_{1},F\right\rangle \longleftrightarrow\left|D_{5},F+1\right\rangle $} \\  & 2. \textcolor{red}{ $\left|D_{2},F\right\rangle \longleftrightarrow\left|D_{5},F+1\right\rangle $} \\ & 3. \textcolor{blue}{ $\left|D_{1},F\right\rangle \longleftrightarrow\left|D_{4},F+1\right\rangle $} \\ & 4. \textcolor{blue}{ $\left|D_{2},F\right\rangle \longleftrightarrow\left|D_{4},F+1\right\rangle $} \\ & 5. \textcolor{blue}{ $\left|D_{4},F\right\rangle \longleftrightarrow\left|D_{5},F+1\right\rangle $} \\ & 6. \textcolor{blue}{ $\left|D_{2},F\right\rangle \longleftrightarrow\left|D_{3},F+1\right\rangle $} \\ & 7. \textcolor{blue}{ $\left|D_{3},F\right\rangle \longleftrightarrow\left|D_{5},F+1\right\rangle $} \\ & 8. \textcolor{blue}{  $\left|D_{1},F\right\rangle \longleftrightarrow\left|D_{3},F+1\right\rangle $} \\ & 9. \textcolor{blue}{ $\left|D_{3},F\right\rangle \longleftrightarrow\left|D_{4},F+1\right\rangle $} \\ & 10. \textcolor{blue}{ $\left|D_{1},F\right\rangle \longleftrightarrow\left|D_{2},F+1\right\rangle $} \\ & 11. \textcolor{red}{ $\left|D_{2},F\right\rangle \longleftrightarrow\left|D_{1},F+1\right\rangle $ }\\ & 12. \textcolor{red}{ $\left|D_{4},F\right\rangle \longleftrightarrow\left|D_{3},F+1\right\rangle $} \\ & 13. \textcolor{red}{ $\left|D_{3},F\right\rangle \longleftrightarrow\left|D_{1},F+1\right\rangle $ }\\ & 14. \textcolor{red}{ $\left|D_{5},F\right\rangle \longleftrightarrow\left|D_{3},F+1\right\rangle $} \\ & 15. \textcolor{red}{ $\left|D_{3},F\right\rangle \longleftrightarrow\left|D_{2},F+1\right\rangle $ }\\ & 16. \textcolor{red}{ $\left|D_{5},F\right\rangle \longleftrightarrow\left|D_{4},F+1\right\rangle $} \\ & 17. \textcolor{red}{ $\left|D_{4},F\right\rangle \longleftrightarrow\left|D_{2},F+1\right\rangle $} \\ & 18. \textcolor{red}{ $\left|D_{4},F\right\rangle \longleftrightarrow\left|D_{1},F+1\right\rangle $} \\ & 19. \textcolor{red}{ $\left|D_{5},F\right\rangle \longleftrightarrow\left|D_{2},F+1\right\rangle $} \\ & 20. \textcolor{red}{ $\left|D_{5},F\right\rangle \longleftrightarrow\left|D_{1},F+1\right\rangle $} \\
		
		\hline \hline
		
	\end{tabular}
	\caption{Transitions between dressed states in Figure~\ref{fig:rabibands} for two-photon, three-photon, and four-photon pumping of a superconducting artificial atom (transmon) in front of a mirror. The transitions between dressed states can be seen in Fig.~1(b) in the main text. The blue (red) color represents amplified (attenuated) Rabi sidebands.
	\label{tab:2}}
\end{table}


\newpage
\vspace*{1cm}

\section{Saturation power of amplification}

To investigate the saturation properties of amplification, we measure the reflection coefficient as a function of probe frequency at $\omega_p$ and probe power at $P_p$. We fixed the pump power $P_{\rm pump}$ and pump frequency $\omega_{\rm pump}$ at the maximum amplification point where the two Rabi sidebands cross, as seen in Fig.~2(c) in the main text, using the parameters of three-photon pumping. We sweep the probe frequency $\omega_p$ around the maximum amplification point on the x axis and sweep the probe power from $P_p = \unit[-161]{dBm}$ to $\unit[-121]{dBm}$ on the y axis. We observe the largest amplification (dark purple color) at weak probe powers; as the probe power increases, the amplification begins to saturate towards unity (light blue color).

\begin{figure}[h!]
	\includegraphics[width=\linewidth]{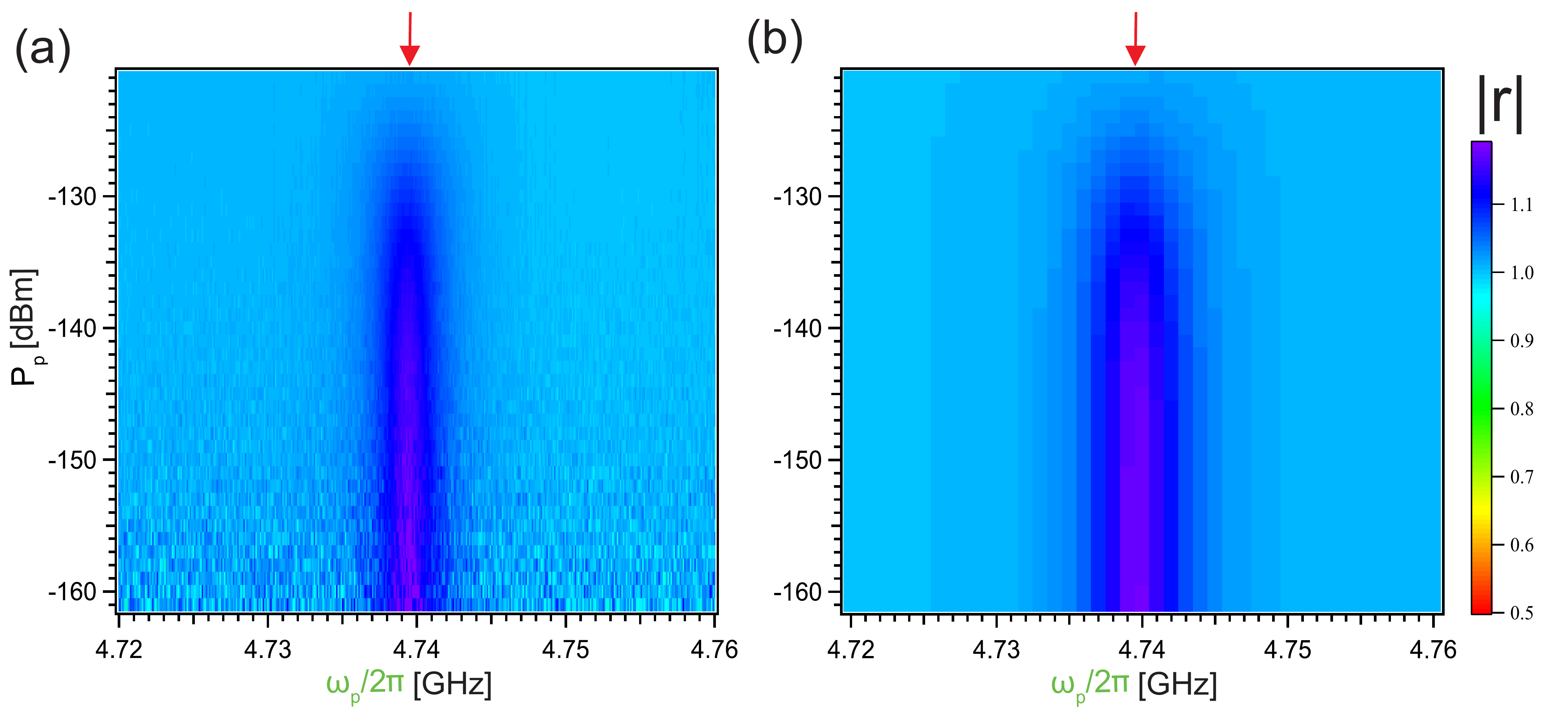}
	\caption{The magnitude $|r|$ of the reflection coefficient for a weak probe with a three-photon pump signal as a function of probe frequency $\omega_p$ and probe power $P_p$. (a) Experimental data and (b) numerical simulation. The linecuts indicated by the red arrows are shown in Fig.~3(a) in the main text. 
	\label{fig:4}}
\end{figure}


\newpage

\section{Noise properties of amplification}

\begin{figure}[h!]
	\includegraphics[width=\linewidth]{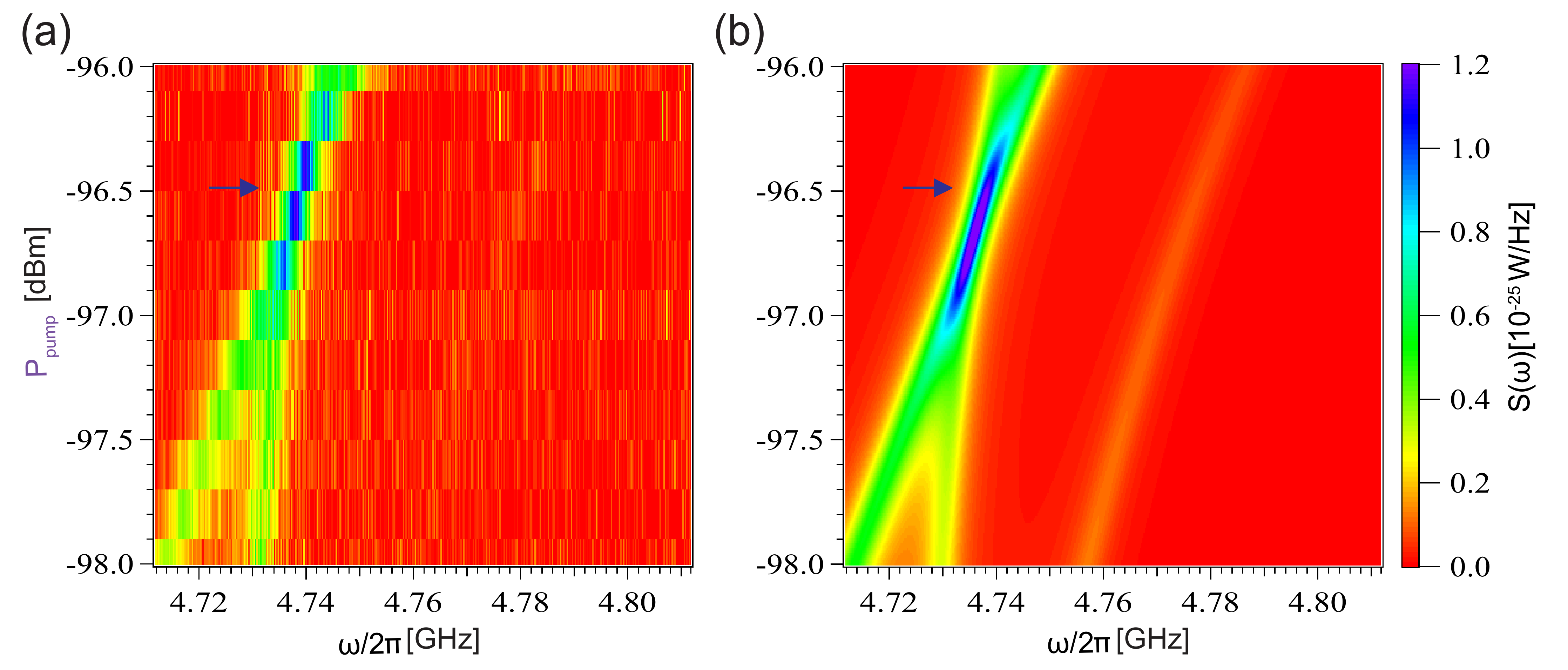}
	\caption{The measured emission spectrum from an artificial atom as a function of frequency $\omega$ and pump power $P_{\rm pump}$. (a) Experimental data and (b) numerical simulation. We first strongly pump the atom with a three-photon pump at frequency $\omega_{\rm pump}$ using an RF source and sweep $P_{\rm pump}$, and then measure the emission using a spectrum analyzer. We extract the emission spectrum by subtracting the pump-off spectral curve from the pump-on spectral curve. The linecuts indicated by the blue arrows are shown in Fig.~3(b) in the main text.
		\label{fig:5}}
\end{figure}

In order to analyze the noise properties of the amplification, we measure the spontaneous emission from the artificial atom. We first strongly pump the atom with a resonant three-photon pump at frequency $\omega_{\rm pump} / 2\pi = \omega_{30} / 6\pi = \unit[4.530]{GHz}$ and sweep the pump power from $P_{\rm pump} = \unit[-98]{dBm}$ to $\unit[-96]{dBm}$ on the y axis, and then measure the spontaneous emission using the experimental setup shown in \figpanel{fig:Device}{b}. Similarly, we measure the spectral curve when the pump is off. Figure~\figpanelNoPrefix{fig:5}{a} shows the extracted emission spectrum by subtracting the pump-off spectral curve from the pump-on spectral curve. Note that we measured this experiment in a new cooldown, and we found that the pump power differs by a few dBm compared to Fig.~2(a) in the main text. When $P_{\rm pump} < \unit[-97.5]{dBm}$, we observe the two Rabi sidebands corresponding to the labels (i) and (ii) in Fig.~2(b) in the main text. As the pump power increases, we observe the same effect: the frequency of these Rabi sidebands (i \& ii) become the same at a certain frequency, where we obtain constructive interference and increased amplification, as can be seen in Fig.~2(b) in the main text. Note that the gain of the HEMT amplifier is \unit[2]{dB} higher at this frequency, due to the frequency dependence of gain. We calibrated the gain at \unit[4.766]{GHz} using the method described in Ref.~\cite{YT2022phaseshaping} to calibrate the gain and attenuation of the line in the setup shown in \figpanel{fig:Device}{b}.

\newpage
\bibliography{supplement.bbl}